\documentclass[journal,twoside,10pt]{IEEEtran}
\usepackage{amsmath}
\usepackage{mathrsfs}

\usepackage{amsfonts}
\usepackage[numbers,sort&compress]{natbib}
\usepackage{graphicx,color,overpic}
\usepackage{amsmath}
\usepackage{times}
\usepackage{latexsym}
\usepackage{bm}
\usepackage{amssymb}
\usepackage{cases}
\usepackage{array}
\usepackage{fancyhdr}
\usepackage{setspace}

\usepackage[caption=false,font=footnotesize,labelfont=rm,textfont=rm]{subfig}
\usepackage{subfig}
\usepackage{url}
\usepackage{algorithm}
\usepackage{algorithmic}
\usepackage{multirow}
\allowdisplaybreaks[4]

\usepackage{citesort}
\usepackage{epsfig}
\usepackage{fancybox}
\usepackage{textcomp}
\usepackage{multirow}
\usepackage{setspace}
\usepackage{psfrag}
\usepackage{makecell}

\newcommand{\qa}{{\bf a}}
\newcommand{\qb}{{\bf b}}
\newcommand{\qd}{{\bf d}}
\newcommand{\qh}{{\bf h}}
\newcommand{\qn}{{\bf n}}
\newcommand{\qv}{{\bf v}}
\newcommand{\qx}{{\bf x}}
\newcommand{\qy}{{\bf y}}

\newcommand{\qA}{{\bf A}}
\newcommand{\qI}{{\bf I}}
\newcommand{\qQ}{{\bf Q}}
\newcommand{\qV}{{\bf V}}

\newcommand{\bbC}{{\mathbb C}}

\begin{document}
\title{Hybrid-Field Channel Tracking for Extremely Large-Scale MIMO Systems with Mobility}

\author{Yilong Liu,~\IEEEmembership{Graduate Student Member,~IEEE},
Xi Yang,~\IEEEmembership{Member,~IEEE},
Ting Liu,~\IEEEmembership{Member,~IEEE},

Yu Han,~\IEEEmembership{Member,~IEEE},
and Shi Jin,~\IEEEmembership{Fellow,~IEEE}
\thanks{Copyright (c) 20xx IEEE. Personal use of this material is permitted. However, permission to use this material for any other purposes must be obtained from the IEEE by sending a request to pubs-permissions@ieee.org.}
\thanks{This work was supported in part by the National Natural Science Foundation of China (NSFC) under Grant U25A20392, Grant 62301221, and Grant 62422105, in part by the Jiangsu Provincial Department of Education Special Project for Central Universities Serving Jiangsu's High-Quality Development: 5G-A/6G Open-source R\&D and Industrial Application under Grant JSE202512310, and in part by the Science and Technology Commission of Shanghai Municipality (STCSM) under Grant 22DZ2229005.
{\it (Corresponding author: Xi Yang.)}}
\thanks{Yilong Liu and Xi Yang are with the School of Information and Electronic Engineering, East China Normal University, Shanghai 200241, China (e-mail: yilongliu@stu.ecnu.edu.cn; xyang@cee.ecnu.edu.cn).}
\thanks{Ting Liu is with the School of Artificial Intelligence, Nanjing University of Information Science and Technology, Nanjing 210044, China (e-mail: liuting@nuist.edu.cn).}
\thanks{Yu Han and Shi Jin are with the National Mobile Communications Research Laboratory, Southeast University, Nanjing 210096, China (e-mail: hanyu@seu.edu.cn; jinshi@seu.edu.cn).}
}
\maketitle

\begin{abstract}
In this paper, we propose a hybrid-field channel tracking algorithm for extremely large-scale multiple-input multiple-output (XL-MIMO) systems with mobility, leveraging the historical channel state information.
The multiple path scenario is considered, and the line-of-sight path is coarsely estimated through the sparse peak search within a narrow window in the fractional Fourier domain based on the temporal continuity of the user's motion.
Moreover, coarse estimations of the non-line-of-sight (NLoS) paths are determined by ensuring the scatterer survival probability, eliminating the necessity of detecting existing NLoS paths.
Then, a Newton-based refinement is applied to the estimated paths before seeking possible new paths.
Numerical results validate that the proposed algorithm achieves superior performance with low computational complexity.
\end{abstract}

\begin{IEEEkeywords}
Channel tracking, XL-MIMO, fractional fourier transform, hybrid-field.
\end{IEEEkeywords}

\section{Introduction}
As one of the promising technologies for sixth-generation ($6\mathrm{G}$) mobile communication networks, extremely large-scale multiple-input multiple-output (XL-MIMO), which deploys extremely large-scale antenna arrays, is expected to unprecedentedly improve the spectral efficiency and spatial resolution, thereby enabling emerging $6\mathrm{G}$ applications \cite{Lu-24COMST}.
In XL-MIMO systems, with the increase of the antenna array size, the Rayleigh distance expands considerably.
Thus, part of the scatterers fall within the near-field region of the antenna array, while the remaining scatterers are still located in the far-field region, leading to the so-called hybrid-field environment \cite{WD-22CL}.

To fully exploit the spatial diversity and multiplexing gains introduced by XL-MIMO, accurate channel state information (CSI) is indispensable, particularly in mobile scenarios.
To alleviate the high pilot overhead for the channel estimation in XL-MIMO systems, recent studies \cite{WD-22CL}, \cite{XYang-24WCL}, \cite{JLu-25TVT} have investigated compressed sensing-based methods in the hybrid-field environment by utilizing the channel sparsity.
For instance, \cite{WD-22CL} proposed a hybrid-field orthogonal matching pursuit (OMP) algorithm that sequentially estimated far-field paths in the angle domain and near-field paths in the polar domain.
A gridless hybrid-field channel estimation algorithm was proposed in \cite{XYang-24WCL} based on the hybrid-field channel sparsity in the fractional Fourier domain.
\cite{JLu-25TVT} developed an adaptive on-grid hybrid-field channel estimation scheme applicable to the scenarios without prior information on the number of propagation paths.

However, all the aforementioned algorithms are mainly designed for static or quasi-static communication environments.
In user mobility scenarios, the CSI varies rapidly, necessitating frequent CSI acquisition and consequently introducing prohibitive computational overhead in XL-MIMO systems.
As a result, existing channel estimation algorithms are extravagant and may even become impractical in such cases \cite{JTan-21JSAC}, highlighting the need for low-complexity alternatives.
To solve this challenge, we propose a channel tracking algorithm for the hybrid-field XL-MIMO system with a moving user, in which the CSI estimated in previous time slots (e.g., historical trajectory) is leveraged to improve estimation accuracy and reduce computational complexity.
Specifically, in XL-MIMO systems, the distance dimension is embedded in the channel vector of near-field paths, so that the CSI inherently contains the angle of arrival (AoA), distance, and velocity information, which enables the construction of the user's historical trajectory.
The primary contributions are summarized as follows.

The discrete fractional Fourier transform (DFrFT) is applied to reveal the hybrid-field channel sparsity.
Then, by leveraging the distinct geometric sensitivities of the line-of-sight (LoS) and the non-LoS (NLoS) paths under user mobility in the hybrid field, we propose a differentiated path-level channel tracking scheme with significantly reduced computational complexity.
Specifically, thanks to the temporal continuity of the user's motion, the LoS path is coarsely estimated based on the sparse peak search within a narrow window determined by the historical CSI in the fractional Fourier domain.
After that, by exploiting the channel tracking interval where the scatterer survival probability remains above a specified threshold, only the path gains of NLoS paths are updated based on the satisfied survival probability threshold.
A Newton-based refinement is also implemented for the estimated paths to enhance the performance before seeking possible new paths.
Numerical results show that the proposed algorithm performs better than existing algorithms with lower computational complexity.

\section{System Model and Problem Formulation}\label{s:Model}
\subsection{System Model}
\begin{figure}[htbp]
       \centering
       \includegraphics[width=0.42\textwidth]{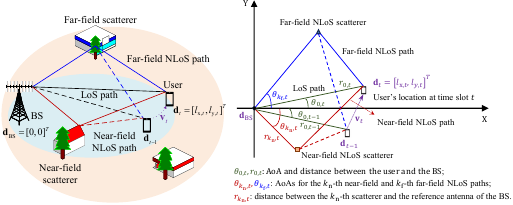}
       \caption{The hybrid-field XL-MIMO system with mobility, where the user is moving from $\qd_{t-1}$ to $\qd_t$ with the velocity of $\qv_t$.}\label{fig:sys_model}
\end{figure}

As illustrated in Fig.\,\ref{fig:sys_model}, we consider an XL-MIMO system, where the base station (BS) equipped with a uniform linear array consisting of $M$ antennas serves a single-antenna user with mobility.
Due to the large aperture size of the antenna array at the BS, the user is located in the hybrid-field scatterer environment.
We adopt a time-slotted communication frame structure, and then, with the uplink pilot $s$ transmitted by the user,\footnote{Without loss of generality, we set $s = 1$.} the received signal of the BS at the $t$-th time slot $\qy_t \in \bbC^{M \times 1}$ can be expressed as
\begin{align}\label{eq:yt}
       \qy_t = \qh_t s + \qn_t,
\end{align}
where $\qn_t$ is the zero-mean complex additive white Gaussian noise and $\qn_t \sim \mathcal{C}\mathcal{N}(\mathbf{0}, \sigma^2_\mathrm{n}\qI_M)$,
$\qh_t \in \bbC^{M \times 1}$ denotes the channel vector at the $t$-th time slot, which can be written by
\begin{align}\label{eq:h_t}
       \qh_t = \qh_{\mathrm{LoS},t} + \qh_{\mathrm{NLoS},t},
\end{align}
with $\qh_{\mathrm{LoS},t}$ and $\qh_{\mathrm{NLoS},t}$ representing the LoS and NLoS components, respectively, as follows\footnote{Due to the movement, the user may be located in the near-field or the far-field of the BS.
For general consideration, we adopt the near-field modeling for the LoS path since the far-field expression can be regarded as a special case of the near-field when the Fresnel expansion is applied.}
\begin{subequations}
\begin{align}
       \qh_{\mathrm{LoS},t} =& \sqrt{\frac{M}{K+1}} \beta_{0,t} e^{j2\pi \omega_{0,t} t} \qb(\theta_{0,t}, r_{0,t}),\\
       \qh_{\mathrm{NLoS},t} =& \sqrt{\frac{M}{K+1}} \left(\sum_{k_\mathrm{f}=1}^{K_\mathrm{f}} \beta_{k_\mathrm{f},t} e^{j2\pi \omega_{k_\mathrm{f},t} t} \qa(\theta_{k_\mathrm{f},t}) \right.\notag\\
       & \left. +\sum_{k_\mathrm{n}=K_\mathrm{f}+1}^{K} \beta_{k_\mathrm{n},t} e^{j2\pi \omega_{k_\mathrm{n},t} t} \qb(\theta_{k_\mathrm{n},t}, r_{k_\mathrm{n},t})\right),
\end{align}
\end{subequations}
where $\beta_{0,t}$, $\beta_{k_\mathrm{f},t}$ and $\beta_{k_\mathrm{n},t}$ denote the complex gain coefficients associated with the LoS path, the far-field and the near-field NLoS paths at the $t$-th time slot, respectively, which explicitly contains the path loss and the phase shift $e^{-j \frac{2\pi \tau_{k,t}}{\lambda}}$ induced by the propagation delay $\tau_{k,t} = \frac{r_{k,t}}{c}$, $\lambda$ is the carrier wavelength, and $c$ is the speed of light.
$K_\mathrm{f}$ and $K_\mathrm{n}$ represent the number of the far-field and the near-field NLoS paths, respectively, and $K = K_\mathrm{f} + K_\mathrm{n}$.
$\theta_{k_\mathrm{f},t}$ and $\theta_{k_\mathrm{n},t}$ are the AoAs for the far-field and the near-field NLoS paths at the $t$-th time slot, respectively.
$\theta_{0,t}$ and $r_{0,t}$ denote the AoA and the distance between the user and the BS at the $t$-th time slot, respectively.
$r_{k_\mathrm{n},t}$ denotes the range for the $k_\mathrm{n}$-th near-field path at the $t$-th time slot, i.e., the distance between the $k_\mathrm{n}$-th scatterer and the reference antenna of the BS.
$\omega_{k,t}$ denotes the Doppler frequency for the $k$-th path at the $t$-th time slot and can be written by
\begin{align}
       \omega_{k,t} = \frac{[\cos \theta_{k,t}, \sin \theta_{k,t}] \cdot \qv_t}{\lambda}, 0 \leq k \leq K,
\end{align}
where $\qv_t \in \bbC^{2 \times 1}$ is the velocity vector of the user at the $t$-th time slot.
$\qa(\theta_{k_\mathrm{f},t})$ and $\qb(\theta_{k_\mathrm{n},t}, r_{k_\mathrm{n},t})$ denote the far-field and near-field steering vectors at the $t$-th time slot, respectively, and we have
\begin{subequations}
\begin{align}
       [\qa(\theta_{k_\mathrm{f},t})]_m =& \frac{1}{\sqrt{M}} e^{-j \frac{2\pi (m-1)d \sin \theta_{k_\mathrm{f},t}}{\lambda}},\\
       [\qb(\theta_{k_\mathrm{n},t}, r_{k_\mathrm{n},t})]_m =& \frac{1}{\sqrt{M}} e^{-j \frac{2\pi}{\lambda}(r_{m-1,k_\mathrm{n},t}-r_{k_\mathrm{n},t})},
\end{align}
\end{subequations}
where $[\cdot]_m$ denotes the $m$-th element of the vector, $d$ is the antenna spacing at the BS, and $r_{m,k_\mathrm{n},t}$ denotes the distance between the $m$-th antenna at the BS and the $k_\mathrm{n}$-th near-field scatterer at the $t$-th time slot, which is given by
\begin{align}\label{eq:r_k}
       r_{m,k_\mathrm{n},t} = \sqrt{r_{k_\mathrm{n},t}^2 + m^2d^2-2mdr_{k_\mathrm{n},t}\sin(\theta_{k_\mathrm{n},t})}.
\end{align}
To characterize the Rician fading propagation environment, we define the Rician factor $\kappa$ as $\kappa = \frac{\|\qh_{\mathrm{LoS},t}\|^2}{\|\qh_{\mathrm{NLoS},t}\|^2}$.

Thanks to the motion characteristics embedded in the historical trajectory, the current user's location can be highly predicted.
Thus, similar to \cite{HXiao-20TCOM}, the widely used Gauss-Markov mobility model is adopted, and then the velocity vector of the user at the $t$-th time slot can be expressed as $\qv_t = \zeta \qv_{t-1} + (1 - \zeta)\bar{\qv} + \sqrt{1 - \zeta^2} \tilde{\qv}$, where $\bar{\qv}$ denotes the asymptotic average velocity vector as $t \rightarrow \infty$, $\zeta \in [0, 1]$ is the randomness parameter,\footnote{Significantly, $\zeta = 0$ indicates that the motion is fully random, i.e., the Brownian motion, $\zeta = 1$ indicates that the motion is linear \cite{RHe-18TWC}.} and $\tilde{\qv} \in \bbC^{2 \times 1}$ is an independent Gaussian random vector, i.e., $[\tilde{\qv}]_i \sim \mathcal{C}\mathcal{N}(\mathbf{0}, \sigma^2_{\mathrm{v},i}), \forall i \in \{1, 2\}$.
As a result, with the BS's location, i.e., $\qd_{\mathrm{BS}} = [0, 0]^T$, the user's location at the $t$-th time slot, i.e., $\qd_t = [l_{\mathrm{x},t}, l_{\mathrm{y},t}]^T$, can be given by
\begin{align}\label{eq:dt}
       \qd_t = \bar{\qd}_t + \tilde{\qd},
\end{align}
where $\bar{\qd}_t = \qd_{t-1} + \qv_{t-1} \Delta_t$, $\Delta_t$ denotes the duration of each time slot, and $\tilde{\qd} \sim \mathcal{N}(0, \sigma^2_{\mathrm{d}}\qI_2)$ represents the motion model error, capturing the deviation from uniform linear motion within each time slot.
Obviously, as shown in Fig.\,\ref{fig:sys_model}, $\theta_{0,t}$ and $r_{0,t}$ should be updated as $\theta_{0,t}=\arctan (\frac{l_{\mathrm{y},t}}{l_{\mathrm{x},t}})$ and $r_{0,t} = ||\qd_t||$, respectively.

\subsection{Problem Formulation}
\begin{figure}[htbp]
       \centering
       \includegraphics[width=0.32\textwidth]{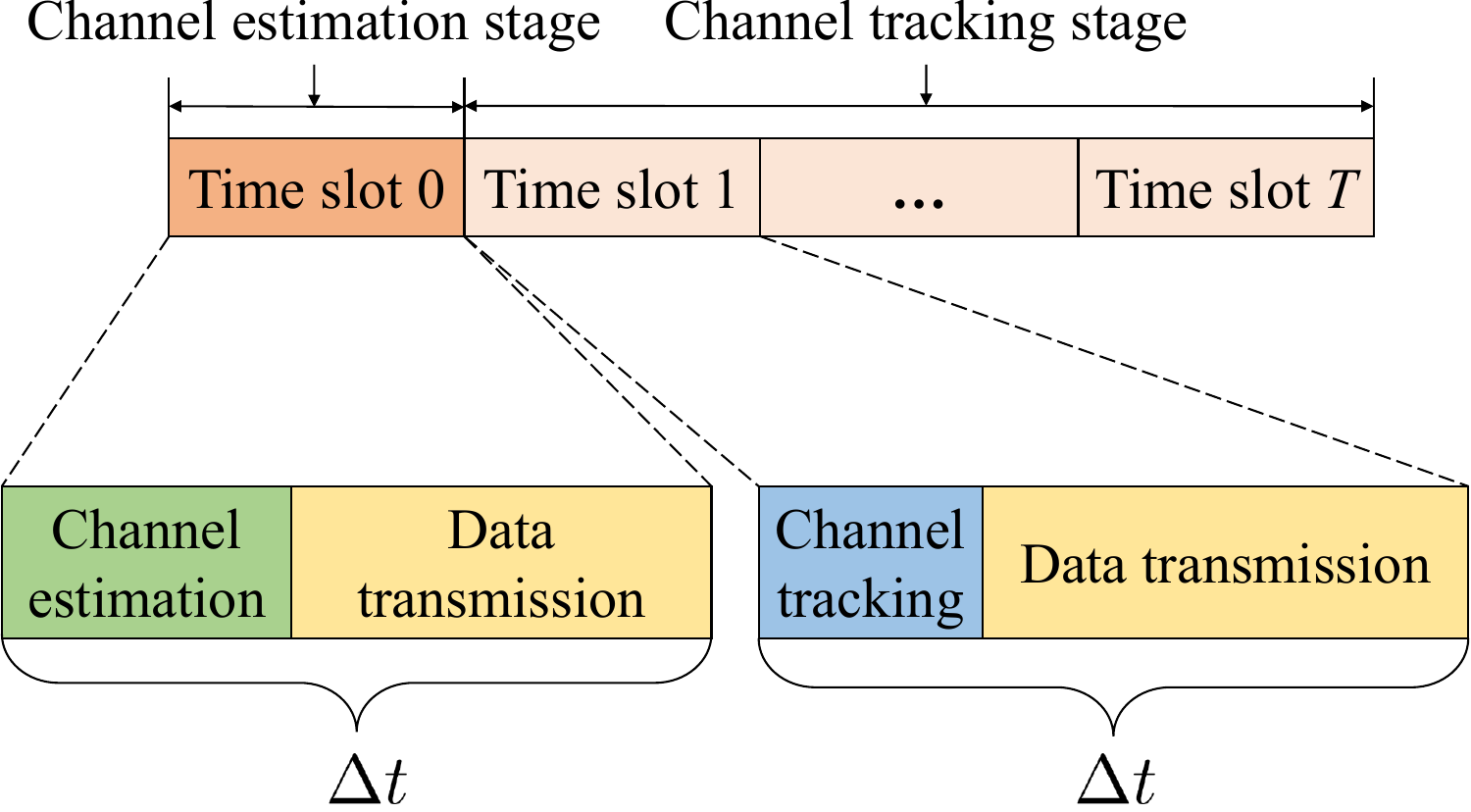}
       \caption{The considered two-stage transmission protocol.}\label{fig:protocol}
\end{figure}

In mobility scenarios where frequent CSI acquisition is necessitated, existing channel estimation algorithms are extravagant due to the prohibitive computational overhead.
Moreover, in hybrid-field scattering environments, the channel sparsity is difficult to exploit directly due to the energy leakage and the energy spread, either in the spatial domain or in the polar domain.
Despite these challenges, the hybrid-field environment also provides some advantages, i.e., $\qh_t$ in \eqref{eq:h_t} contains the user's location $\qd_{t-1}$ and the user's velocity $\qv_{t-1}$.
This allows the historical CSI to be exploited for channel tracking in XL-MIMO systems with mobility, thereby reducing estimation overhead and enhancing estimation accuracy.
Therefore, as illustrated in Fig.\,\ref{fig:protocol}, we consider a two-stage transmission protocol, which is composed of the channel estimation stage (time slot $0$) and the channel tracking stage (time slot $1$-$T$).
In this paper, we focus on the channel tracking process, where the CSI at each time slot is estimated by leveraging the CSI estimated in previous time slots, and the initial channel estimation process at time slot $0$ can be performed based on \cite{XYang-24WCL}.
Our objective is to estimate the hybrid-field channel at the $t$-th time slot $\qh_t$ from the received signal $\qy_t$ in \eqref{eq:yt} with obtained $\qv_{t-1}$, $\qd_{t-1}$, and $\qh_{t-1}$ for $t > 0$.\footnote{Note that these historical CSI can be imperfect, since the proposed three-sigma rule-based search window in Section\,\ref{subs:coarse_est} tolerates these estimation errors.}

\section{Hybrid-Field Channel Tracking}\label{s:Tracking}
In this section, we investigate the channel tracking in hybrid-field environments and propose a hybrid-field channel tracking algorithm, consisting of the coarse estimation, the Newton-based refinement, and the new paths detection steps, in which the DFrFT is employed to excavate the channel sparsity in hybrid-field environments.

\subsection{Discrete Fractional Fourier Transform}
To utilize the inherent sparsity of the hybrid-field channel in the fractional Fourier domain, where different propagation paths manifest as distinct peaks in the DFrFT spectra, we apply the DFrFT to the received signal.
Firstly, by employing the Fresnel expansion, $r_{m,k_\mathrm{n},t}$ in \eqref{eq:r_k} can be approximated by
\begin{align}
       r_{m,k_\mathrm{n},t} \approx r_{k_\mathrm{n},t} - md \sin(\theta_{k_\mathrm{n},t}) + \frac{m^2d^2\cos^2(\theta_{k_\mathrm{n},t})}{2r_{k_\mathrm{n},t}},
\end{align}
and thus the received signal $\qy_t$ in \eqref{eq:yt} can be recast as
\begin{align}
       \qy_t = \sum_{k = 0}^{K} \gamma_{k,t} e^{j2\pi \omega_{k,t} t} \tilde{\qb}(\phi_{k,t}, \varphi_{k,t}) + \tilde{\qn}_t,
\end{align}
where $\tilde{\qn}_t$ denotes the effective noise consisting of the approximation error and $\qn_t$, $\gamma_{k,t} \triangleq \sqrt{\frac{M}{K+1}} \beta_{k,t}$ is the normalized complex path gain,
and $\tilde{\qb}(\phi_{k,t}, \varphi_{k,t})$ represents the normalized steering vector, which can be expressed as
\begin{align}
      [\tilde{\qb}(\phi_{k,t}, \varphi_{k,t})]_m = \frac{1}{\sqrt{M}} e^{j (m-1)\phi_{k,t}}e^{-j (m-1)^2\varphi_{k,t}},
\end{align}
where $\phi_{k,t} \triangleq \frac{2\pi d \sin(\theta_{k,t})}{\lambda}$ and $\varphi_{k,t} \triangleq \frac{\pi d^2 \cos^2(\theta_{k,t})}{\lambda r_{k,t}}$.
For the far-field paths, we have $\qa(\theta_{k_\mathrm{f},t}) = \tilde{\qb}(-\phi_{k,t}, 0)$.

Then, the $p$-th order DFrFT of $\mathbf{y}_t$ is given by \cite{XYang-24WCL}
\begin{align}
       \mathcal{F}^p \{\qy_t\} = c_{\alpha_t} \sum_{m=0}^{M-1} e^{\frac{j\pi \csc \alpha_t}{M}(q_t^2\cos \alpha_t -2q_tm+m^2\cos \alpha_t)}[\qy_t]_m,
\end{align}
where $\alpha_t$ denotes the rotation angle of the spatial-angle plane at the $t$-th time slot as the DFrFT is interpreted as a rotation in the time-frequency plane, $q \in \{0, \dots, M-1\}$ denotes the corresponding DFrFT index for each $\alpha_t$, and $c_{\alpha_t} \triangleq \sqrt{(1-j\cot \alpha_t)/M}$.
By exploiting the order additivity property of the DFrFT \cite{Ozaktas-96TSP}, we define $\alpha_t \triangleq p\frac{\pi}{2}$ with $p \in [0.5, 1.5]$.
Note that the DFrFT reduces to the conventional Fourier transform when $p = 1$, and the DFrFT of $\qy_t$ can be efficiently computed based on the fast Fourier transform-based convolution, resulting in an overall computational complexity of $\mathcal{O}(M\log M)$.
Furthermore, each propagation path in the DFrFT domain is jointly determined by the rotation angle $\alpha_t$ and the frequency index $q$, corresponding to the distance and angle parameters, respectively.
As a result, the estimation of path parameters $(r_{k,t}, \theta_{k,t})$ (or $(\phi_{k,t}, \varphi_{k,t})$) has been transformed into the sparse peak search in the DFrFT spectra.

\subsection{Coarse Estimation}\label{subs:coarse_est}
In this step, due to the temporal continuity of the user's motion, the LoS path only needs to be detected within a narrow two-dimensional search window when the historical CSI is available, whose center is determined by the user's location in the last time slot, thereby significantly reducing the search range and complexity.
For the NLoS paths, since most scatterers remain valid between adjacent time slots when the survival probability threshold is satisfied, their large-scale fading parameters nearly remain stable.
Consequently, the re-detection for the path parameters of AoAs and distances is unnecessary, and only their path gains need to be updated, further reducing the computational complexity.

\subsubsection{Coarse Estimation for the LoS Path}\label{subs:coarse_est_LoS}
With the obtained $\qv_{t-1}$ and $\qd_{t-1}$ in the last time slot, $\qd_t$ can be predicted as $\qd_t \sim \mathcal{N}(\bar{\qd}_t, \sigma^2_{\mathrm{d}}\qI_2)$ via \eqref{eq:dt}.
Then, with $\bar{\qd}_t = [\bar{l}_{\mathrm{x},t}, \bar{l}_{\mathrm{y},t}]^T$, the distribution of $r_{0,t}$ and $\theta_{0,t}$ are derived based on the delta method \cite{Oehlert-92}, as follows
\begin{subequations}\label{eq:r_theta_distribution}
\begin{align}
       r_{0,t} &\overset{\mathrm{approx}}{\sim} \mathcal{N}(||\bar{\qd}_t||, \sigma^2_{\mathrm{d}}), \\
       \theta_{0,t} &\overset{\mathrm{approx}}{\sim} \mathcal{N} \left(\arctan \left(\frac{\bar{l}{\mathrm{y},t}}{\bar{l}_{\mathrm{x},t}} \right), \frac{\sigma^2_{\mathrm{d}}}{||\bar{\qd}_t||^2} \right).
\end{align}
\end{subequations}

To further mitigate the Gaussian random disturbance in $r_{0,t}$ and $\theta_{0,t}$ as shown in \eqref{eq:r_theta_distribution} and enhance user localization accuracy, we propose to perform peak detection on the DFrFT spectrum.
Note that the computational complexity has been significantly reduced since we constrain the search range within a narrow search window determined by the three-sigma rule based on the above derived approximate distributions of $r_{0,t}$ and $\theta_{0,t}$, i.e., $p \in [p_{\mathrm{low}}, p_{\mathrm{high}}]$ with $p_i = \frac{2}{\pi} \arctan ( \frac{\lambda r_{i,t}}{M \nu d^2 \cos^2 \theta_{i,t}} ), \forall i \in \{\mathrm{low}, \mathrm{high}\}$,
where $r_{\mathrm{low},t} = ||\bar{\qd}_t|| - 3\sigma_{\mathrm{d}}$, $r_{\mathrm{high},t} = ||\bar{\qd}_t|| + 3\sigma_{\mathrm{d}}$, $\theta_{\mathrm{low},t} = \arctan(\frac{\bar{l}_{\mathrm{y},t}}{\bar{l}_{\mathrm{x},t}}) - \frac{3\sigma_{\mathrm{d}}}{||\bar{\qd}_t||}$, and $\theta_{\mathrm{high},t} = \arctan(\frac{\bar{l}_{\mathrm{y},t}}{\bar{l}_{\mathrm{x},t}}) + \frac{3\sigma_{\mathrm{d}}}{||\bar{\qd}_t||}$.
By searching the peak with maximum amplitude from the DFrFT spectra of the residual vector at the $t$-th time slot, i.e., $\qy_{\mathrm{r}, t}$, the coarse estimates of $\phi_{0,t}$ and $\varphi_{0,t}$ can be given by
\begin{subequations}\label{eq:coarse_est_AoA}
\begin{align}
       \hat{\phi}_{0,t} =& \frac{2\pi \bar{q}_t \csc (p_{t,\bar{i}}\pi/2)}{M \nu},\\
       \hat{\varphi}_{0,t} =& \frac{\pi \cot (p_{t,\bar{i}}\pi/2)}{M \nu},
\end{align}
\end{subequations}
where $\nu$ is the oversampling factor, $\bar{i}$ and $\bar{q}_t$ are determined by $(\bar{i}, \bar{q}_t) = \arg \underset{i, q_t}{\max} [\qV]_{i, q_t}$ with $\qV \triangleq [\mathcal{F}^{p_{\mathrm{low}}} \{\qy_{\mathrm{r}, t}\},$ $\dots, \mathcal{F}^{p_{i}} \{\qy_{\mathrm{r}, t}\}, \dots, \mathcal{F}^{p_{\mathrm{high}}} \{\qy_{\mathrm{r}, t}\}]$,
where we have $p_{i} = p_{\mathrm{low}} + \frac{(i-1)(p_{\mathrm{high}}-p_{\mathrm{low}})}{P-1}, \forall i \in \{1, 2, \dots, P\}$, and $P$ is the number of $p_{i}$.
Note that we initialize $\qy_{\mathrm{r},t}$ as $\qy_{\mathrm{r},t} = [\qy_t^T, \mathbf{0}_{M(\nu-1)}^T]^T$.
Then, $\gamma_{0,t}$ can be estimated as
\begin{align}\label{eq:coarse_est_gamma}
       \hat{\gamma}_{0,t} = \frac{\{e^{j2\pi \hat{\omega}_t t}\tilde{\qb}(\hat{\phi}_{0,t}, \hat{\varphi}_{0,t})\}^H \tilde{\qy}_{\mathrm{r}, t}}{\|e^{j2\pi \hat{\omega}_t t}\tilde{\qb}(\hat{\phi}_{0,t}, \hat{\varphi}_{0,t})\|^2},
\end{align}
where $\tilde{\qy}_{\mathrm{r}, t}$ is of the first $M$ elements of $\qy_{\mathrm{r},t}$, and
\begin{align}\label{eq:omega_hat}
       \hat{\omega}_t = \frac{[\cos \hat{\theta}_t, \sin \hat{\theta}_t]\cdot \{\zeta \qv_{t-1} + (1 - \zeta)\bar{\qv}\}}{\lambda}.
\end{align}
Then, we update $\qy_{\mathrm{r},t}$ as
\begin{align}\label{eq:yr_dot}
       \dot{\qy}_{\mathrm{r},t} = \qy_{\mathrm{r},t} - \hat{\gamma}_{0,t}\bar{\qb}(\hat{\phi}_{0,t}, \hat{\varphi}_{0,t}).
\end{align}
where $\bar{\qb}(\hat{\phi}_t, \hat{\varphi}_t) \triangleq [e^{j2\pi \hat{\omega}_t t}\tilde{\qb}^T(\hat{\phi}_t, \hat{\varphi}_t), \mathbf{0}_{M(\nu-1)}^T ]^T$.

However, the DFrFT-based peak search relies on discrete grids, whereas the path parameters are continuously distributed, leading to the off-the-grid effect.
Consequently, the estimation accuracy is limited, which will be improved by the Newton-based refinement in Section\,\ref{ssec:refinement}.

\subsubsection{Coarse Estimation for NLoS Paths}
Since each NLoS scatterer undergoes a birth-death process over time, the survival probability of any arbitrary NLoS scatterer after time $\Delta t$ can be computed as $P_{\mathrm{T}}(\Delta t) = e^{-\lambda_{\mathrm{R}} \frac{P_{\mathrm{F}}(||\Delta \qv||)\Delta t}{D}}$ \cite{SWu-18TCOMM}, where $\lambda_{\mathrm{R}}$ denotes the recombination rate of the scatterer, $P_{\mathrm{F}}$ denotes the percentage of moving scatterers, $\Delta \qv$ is the mean relative velocity between the user and the scatterer, and $D$ is a scenario-dependent coefficient describing the space correlation.
Therefore, most scatterers can survive across two consecutive channel tracking cycles by properly adjusting the channel tracking interval $\Delta t$, i.e., $ P_{\mathrm{T}}(\Delta t) \geq \xi$, with $\xi$ denoting the required survival probability threshold.
In this case, the geometry of NLoS paths characterized by the steering vector remains essentially unchanged between two consecutive channel tracking cycles, whereas the complex path gains may vary significantly.
As a result, the estimates of path parameters from the last time slot provide good approximations of their current values, and then we obtain the coarse estimates $\hat{\phi}_{k,t} = \hat{\phi}_{k,t-1}$ and $\hat{\varphi}_{k,t} = \hat{\varphi}_{k,t-1}$ for $1 \leq k \leq K$ when $\Delta t \leq \Delta t_{\xi}$, where $\Delta t_{\xi}$ represents the maximum channel tracking interval, and is calculated as
\begin{align}\label{eq:delta_t}
       \Delta t_{\xi} = - \frac{D \ln(\xi)}{\lambda_{\mathrm{R}} P_{\mathrm{F}} (||\Delta \qv||)}.
\end{align}
Accordingly, $\gamma_{k,t}$ can be estimated as
\begin{align}\label{eq:coarse_est_gamma_k}
       \hat{\gamma}_{k,t} = \frac{\{\tilde{\qb}(\hat{\phi}_{k,t}, \hat{\varphi}_{k,t})\}^H \dot{\qy}_{\mathrm{r}, t}}{\| \tilde{\qb}(\hat{\phi}_{k,t}, \hat{\varphi}_{k,t})\|^2},
\end{align}
where $\hat{\omega}_{k,t}$ is obtained by substituting $(\hat{\phi}_{k,t}, \hat{\varphi}_{k,t})$ into \eqref{eq:omega_hat}, and we update
\begin{align}\label{eq:yr_dot_NLoS}
       \dot{\qy}_{\mathrm{r},t} = \dot{\qy}_{\mathrm{r},t} - \sum_{k=1}^{K}\hat{\gamma}_{k,t}\bar{\qb}(\hat{\phi}_{k,t}, \hat{\varphi}_{k,t}).
\end{align}

\begin{figure}[htbp]
       \centering
       \includegraphics[width=0.3\textwidth]{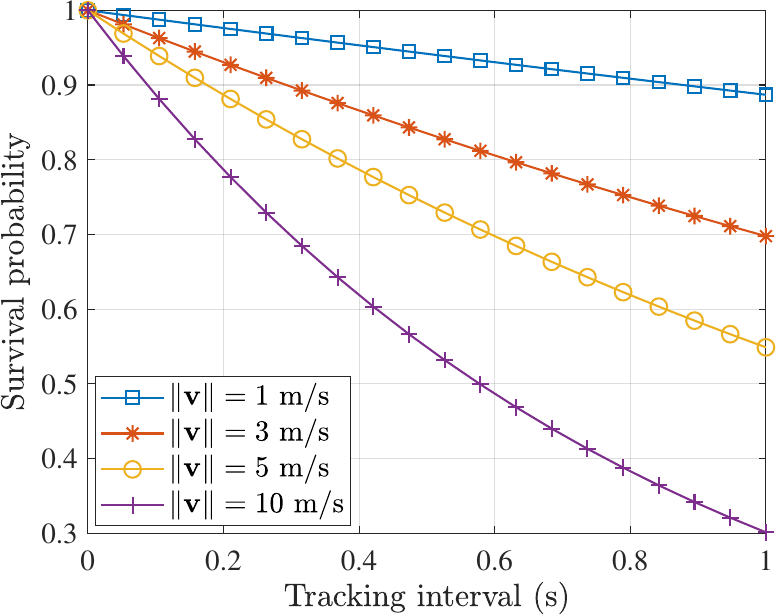}
       \caption{Survival probability versus tracking interval with different velocities for the user.}\label{fig:Survival_probability}
\end{figure}

To investigate the value of $\Delta t$, we depict the survival probability $P_{\mathrm{T}}(\Delta t)$ in Fig.\,\ref{fig:Survival_probability} with different velocities for the user, in which we consider $\lambda_{\mathrm{R}} = 4$, $P_{\mathrm{F}} = 0.3$, and $D = 10\, \mathrm{m}$ \cite{3GPP-38.901}.
With $\xi = 0.8$, the maximum tracking intervals in low- and medium-mobility scenarios (e.g., indoor users, pedestrians, cyclists, and robots) can be more than $0.38\, \mathrm{s}$, and thus exceed multiple radio frames (10 ms in 5G new radio), demonstrating the feasibility of the proposed algorithm.

\subsection{Refinement Based on Newton Method}\label{ssec:refinement}
To alleviate the off-the-grid effect of the LoS path and further improve the estimation accuracy of the NLoS paths, we propose to apply the Newton-based refinements to these obtained coarse estimations in this subsection.

\subsubsection{Local Refinement}\label{sssec:local_refinement}
For each obtained $(\hat{\phi}_{k,t}, \hat{\varphi}_{k,t})$ in $\mathcal{P} = \{(\hat{\phi}_{0,t}, \hat{\varphi}_{0,t}), \dots, (\hat{\phi}_{K,t}, \hat{\varphi}_{K,t})\}$ and the corresponding $\hat{\gamma}_{k,t}$ in $\mathcal{Q} = \{\hat{\gamma}_{0,t}, \dots, \hat{\gamma}_{K,t}\}$, we perform $R_\mathrm{L}$ times of local refinements at this step.
Note that the principle of the refinement is to minimize the residual energy $E(\phi_{k,t}, \varphi_{k,t}) = \| \dot{\qy}_{\mathrm{r},t} - \hat{\gamma}_{k,t}\bar{\qb}(\phi_{k,t}, \varphi_{k,t}) \|^2$, for $(j+1)$-th iteration, we have
\begin{subequations}\label{eq:refine}
\begin{align}
       &\hat{\gamma}_{k,t}^{(j+1)} = \frac{\{\bar{\qb}(\hat{\phi}_{k,t}^{(j)}, \hat{\varphi}_{k,t}^{(j)})\}^H}{\|\bar{\qb}(\hat{\phi}_{k,t}^{(j)}, \hat{\varphi}_{k,t}^{(j)})\|^2} \left( \dot{\qy}_{\mathrm{r},t}^{(j)} + \hat{\gamma}_{k,t}^{(j)} \bar{\qb}(\hat{\phi}_{k,t}^{(j)}, \hat{\varphi}_{k,t}^{(j)})\right),\label{eq:refine_gamma}\\
       &\dot{\qy}_{\mathrm{r},t}^{(j+1)} = \dot{\qy}_{\mathrm{r},t}^{(j)} + \left( \hat{\gamma}_{k,t}^{(j)} - \hat{\gamma}_{k,t}^{(j+1)}\right)\bar{\qb}(\hat{\phi}_{k,t}^{(j)}, \hat{\varphi}_{k,t}^{(j)}),\\
       &[\hat{\phi}_{k,t}^{(j+1)}, \hat{\varphi}_{k,t}^{(j+1)}]^T = [\hat{\phi}_{k,t}^{(j)}, \hat{\varphi}_{k,t}^{(j)}]^T - \qQ^{-1}\qx,\label{eq:refine_phi}
\end{align}
\end{subequations}
where $\qQ \in \mathbb{C}^{2 \times 2}$ denotes the Hessian matrix for $E(\phi_{k,t}, \varphi_{k,t})$, and $\qx \in \mathbb{C}^{2 \times 1}$ is the gradient vector.
Significantly, the updates of $(\hat{\phi}_{k,t}^{(j+1)}, \hat{\varphi}_{k,t}^{(j+1)})$ are only adopted when the Hessian matrix is negative definite and the residual energy decreases.
These conditions ensure that each refinement iteration effectively reduces the residual energy, guaranteeing its convergence.

\subsubsection{Global Refinement}
Each obtained $(\hat{\phi}_{k,t}, \hat{\varphi}_{k,t})$ in $\mathcal{P}$ and the corresponding $\hat{\gamma}_{k,t}$ in $\mathcal{Q}$ are iteratively refined by implementing local refinement for $R_\mathrm{G}$ times at this step.
Subsequently, to further reduce the energy of the residual vector $\dot{\qy}_{\mathrm{r},t}$, $\hat{\boldsymbol{\gamma}}_t = [\hat{\gamma}_{0,t}, \dots, \hat{\gamma}_{K,t}]^T$ should be updated based on $\mathcal{P}$ and the least-square method, which can be expressed as
\begin{align}\label{eq:glo_refine_gamma}
       \hat{\boldsymbol{\gamma}}_t = (\qA^H\qA + \sigma_0^2\qI)^{-1}\qA^H\qy_t,
\end{align}
where $\qA = [\tilde{\qb}(\hat{\phi}_{0,t}, \hat{\varphi}_{0,t}), \dots, \tilde{\qb}(\hat{\phi}_{K,t}, \hat{\varphi}_{K,t})]$, $\sigma_0^2\qI$ is the regularization term that ensures the full rank of $\qA^H\qA + \sigma_0^2\qI$, and we set $\sigma_0^2 = 10^{-5}$ in the simulation.
Then, we update
\begin{align}\label{eq:glo_refine_yr}
       \dot{\qy}_{\mathrm{r},t} = \qy_t - \qA \hat{\boldsymbol{\gamma}}_t.
\end{align}
The Newton-based refinement algorithm can be formulated in \textbf{Algorithm\,\ref{alg:refinement}}.

\begin{algorithm}[!ht]
       \small
       \caption{Newton-Based Refinement Algorithm}
       \begin{algorithmic} [1]\label{alg:refinement}
       \FOR{each $(\hat{\phi}_{k,t}, \hat{\varphi}_{k,t}) \in \mathcal{P}$ and $\hat{\gamma}_{k,t} \in \mathcal{Q}$}
       \STATE Refine $(\hat{\phi}_{k,t}, \hat{\varphi}_{k,t})$ and $\hat{\gamma}_{k,t}$ by $R_\mathrm{L}$ times via \eqref{eq:refine}.
       \FOR{$i = 1, \dots, R_\mathrm{G}$}
       \FOR{each $(\hat{\phi}_{k,t}, \hat{\varphi}_{k,t}) \in \mathcal{P}$ and $\hat{\gamma}_{k,t} \in \mathcal{Q}$}
       \STATE Refine $(\hat{\phi}_{k,t}, \hat{\varphi}_{k,t})$ and $\hat{\gamma}_{k,t}$ by $R_\mathrm{L}$ times via \eqref{eq:refine}.
       \ENDFOR
       \ENDFOR
       \STATE Update $\hat{\boldsymbol{\gamma}}_t$ and $\dot{\qy}_{\mathrm{r},t}$ according to \eqref{eq:glo_refine_gamma} and \eqref{eq:glo_refine_yr}.
       \ENDFOR
\end{algorithmic}
\end{algorithm}

\subsection{Detect New Paths}
After the refinements for existing paths obtained by coarse estimations, we seek for possible new paths based on the peak detection step in Section \ref{subs:coarse_est}, in which the search window is set as $p \in [0.5, 1.5]$ and the number of samples for $p$ is denoted as $\dot{P}$, thus we obtain $(\hat{\phi}_{\mathrm{new},t}, \hat{\varphi}_{\mathrm{new},t})$ and $\hat{\gamma}_{\mathrm{new},t}$ via \eqref{eq:coarse_est_AoA} and \eqref{eq:coarse_est_gamma}.
Accordingly, we update $\mathcal{P} = \{\mathcal{P}, (\hat{\phi}_{\mathrm{new},t}, \hat{\varphi}_{\mathrm{new},t})\}$ and $\mathcal{Q} = \{\mathcal{Q}, \hat{\gamma}_{\mathrm{new},t}\}$, and $\dot{\qy}_{\mathrm{r},t}$ is updated as
\begin{align}\label{eq:yr_dot_new}
       \dot{\qy}_{\mathrm{r},t} = \dot{\qy}_{\mathrm{r},t} - \hat{\gamma}_{\mathrm{new},t}\bar{\qb}(\hat{\phi}_{\mathrm{new},t}, \hat{\varphi}_{\mathrm{new},t}).
\end{align}
Then, all the obtained $(\hat{\phi}_{\mathrm{new},t}, \hat{\varphi}_{\mathrm{new},t})$ in $\mathcal{P}$ and $\hat{\gamma}_{\mathrm{new},t}$ in $\mathcal{Q}$ should be refined based on \textbf{Algorithm\,\ref{alg:refinement}}.

\subsection{Overall Channel Tracking Algorithm}
The channel tracking algorithm terminates when the residual energy $\|\dot{\qy}_{\mathrm{r},t}\|^2$ is less than the threshold $\tau$, which is set as $\tau = \frac{M}{\rho}$ with $\rho$ denoting the signal-to-noise ratio (SNR), and the hybrid-field channel is estimated as
\begin{align}\label{eq:h_hat}
       \hat{\qh}_t = \sum_{k = 0}^{\hat{K}} \hat{\gamma}_{k,t} e^{j2\pi \hat{\omega}_{k,t} t} \tilde{\qb}(\hat{\phi}_{k,t}, \hat{\varphi}_{k,t}),
\end{align}
where $\hat{K} = \mathrm{dim}(\mathcal{P})$ is the total number of the detected paths, and $\mathrm{dim}(\cdot)$ denotes the dimension of a subspace (or vector space).
The overall hybrid-field channel tracking algorithm is summarized in \textbf{Algorithm\,\ref{alg:tracking}}.

\begin{algorithm}[htbp]
       \small
       \caption{Hybrid-Field Channel Tracking Algorithm}
       \begin{algorithmic} [1]\label{alg:tracking}
       \STATE \textbf{Initialize:} $\qy_{\mathrm{r},t} = [\qy_t^T, \mathbf{0}_{M(\nu-1)}^T]^T$, $\dot{\qy}_{\mathrm{r},t} = \qy_{\mathrm{r},t}$.
       \STATE Obtain $(\hat{\phi}_{0,t}, \hat{\varphi}_{0,t})$ and $\hat{\gamma}_{0,t}$ based on \eqref{eq:coarse_est_AoA} and \eqref{eq:coarse_est_gamma}, and update $\dot{\qy}_{\mathrm{r},t}$ via \eqref{eq:yr_dot}.
       \STATE Determine $\Delta t$ based on \eqref{eq:delta_t}, obtain $\hat{\phi}_{k,t} = \hat{\phi}_{k,t-1}$, $\hat{\varphi}_{k,t} = \hat{\varphi}_{k,t-1}$, and $\hat{\gamma}_{k,t}$ via \eqref{eq:coarse_est_gamma_k}, and update $\dot{\qy}_{\mathrm{r},t}$ via \eqref{eq:yr_dot_NLoS}.
       \STATE Refine all $(\hat{\phi}_{k,t}, \hat{\varphi}_{k,t}) \in \mathcal{P}$ and $\hat{\gamma}_{k,t} \in \mathcal{Q}$ based on \textbf{Algorithm\,\ref{alg:refinement}}.
       \WHILE{$\|\dot{\qy}_{\mathrm{r},t}\|^2 \geq \tau$}
       \STATE Obtain $(\hat{\phi}_{\mathrm{new},t}, \hat{\varphi}_{\mathrm{new},t})$ and $\hat{\gamma}_{\mathrm{new},t}$ based on \eqref{eq:coarse_est_AoA} and \eqref{eq:coarse_est_gamma}, and update $\dot{\qy}_{\mathrm{r},t}$ via \eqref{eq:yr_dot_new}.
       \STATE Update $\mathcal{P} = \{\mathcal{P}, (\hat{\phi}_{\mathrm{new},t}, \hat{\varphi}_{\mathrm{new},t})\}$, $\mathcal{Q} = \{\mathcal{Q}, \hat{\gamma}_{\mathrm{new},t}\}$.
       \STATE Refine $(\hat{\phi}_{\mathrm{new},t}, \hat{\varphi}_{\mathrm{new},t}) \in \mathcal{P}$ and $\hat{\gamma}_{\mathrm{new},t} \in \mathcal{Q}$ based on \textbf{Algorithm\,\ref{alg:refinement}}.
       \ENDWHILE
       \STATE Obtain $\hat{\qh}_t$ based on \eqref{eq:h_hat}.
       \end{algorithmic}
\end{algorithm}

\begin{table}[!htbp]
       \footnotesize
       \centering
       \caption{Comparison of Computational Complexity}\label{table:Complexity}
       \begin{tabular}{|c|c|}
       \hline
       Algorithm              & Computational Complexity   \\ \cline{1-2}
       Proposed algorithm     & $\!\!\!$\makecell{$\mathcal{O}\{(\hat{K}-K)\dot{P}M \nu\log(M \nu)+ PM \nu\log(M \nu)$ \\ $ + MKR_\mathrm{L}R_\mathrm{G}+ M\hat{K}(\hat{K}-K)R_\mathrm{L}R_\mathrm{G}+ MK^2$\\$+K^3 + (\hat{K}-K)(M\hat{K}^2+\hat{K}^3) \}$}$\!\!\!$ \\ \cline{1-2}
       Hybrid-field OMP \cite{WD-22CL}      & $\mathcal{O}\{M\nu\frac{K^3}{8}+\frac{K^3}{8}S+MS\}$    \\ \cline{1-2}
       $\!\!\!$Hybrid-field NOMP \cite{XYang-24WCL}$\!\!\!$    & \makecell{$\mathcal{O}\{\hat{K}PM \nu\log(M \nu)$\\$ + M\hat{K}^2R_\mathrm{L}R_\mathrm{G} + \hat{K}^4 + M\hat{K}^3\}$}   \\ \cline{1-2}
       Near-field OMP \cite{CD-22TCOMM}        & $\mathcal{O}\{K^3S+MS\}$  \\ \hline
       \end{tabular}
\end{table}

The computational complexities of \textbf{Algorithm\,\ref{alg:tracking}} and baseline algorithms are calculated in Table\,\ref{table:Complexity}, where $S$ denotes the number of sampled grids.
Then, we discuss the convergence of \textbf{Algorithm\,\ref{alg:tracking}}.
The convergence of local refinement has been established in Section \ref{ssec:refinement}, and the refinement of $\hat{\boldsymbol{\gamma}}_t$ based on the least-square method ensures a reduction in the residual energy.
Thus, the convergence of \textbf{Algorithm\,\ref{alg:tracking}} is guaranteed.

\section{Numerical Results}\label{s:Simulations}
In this section, we conduct simulations to validate the effectiveness and superiority of the channel tracking algorithm.
The normalized mean square error (NMSE) is employed to evaluate the channel estimation performance, which is defined as $\mathrm{NMSE} = \mathbb{E}(\frac{\|\hat{\qh}_t-\qh_t\|^2}{\|\qh_t\|^2})$.
The considered XL-MIMO system operates at $28\,\mathrm{GHz}$, and the corresponding wavelength is $\lambda =0.0107\,\mathrm{m}$.
We assume that the location and the velocity of the user at the $0$-th time slot are $\qd_0 = [57\,\mathrm{m}, 0\,\mathrm{m}]^T$ and $\qv_0 = [\frac{\sqrt{2}}{2}\,\mathrm{m/s}, \frac{\sqrt{2}}{2}\,\mathrm{m/s}]^T$, respectively.
Unless otherwise specified, we perform $2000$ channel realizations at each SNR for the NMSE, and we set $M=256$, $d=\frac{\lambda}{4}$, $\nu = 4$, $\xi = 0.9$, $\sigma^2_{\mathrm{v},i} = 1$ for $i \in \{1,2\}$, $\sigma^2_{\mathrm{d}} = 10$, $R_\mathrm{L} = R_\mathrm{G} = 5$, $\kappa=8$, $\theta_{k,t} \sim \mathcal{U}(-\frac{\pi}{3},\frac{\pi}{3})$, and $r_{k,t} \sim \mathcal{U}(3.4\,\mathrm{m}, 57.9\,\mathrm{m})$, where $\mathcal{U}(a,b)$ represents the uniform distribution over the interval $[a, b]$.
Additionally, the initial numbers of near-field and far-field paths at time slot $t-1$ are set as $K_\mathrm{n} = K_\mathrm{f} =5$.

\begin{figure}[htbp]
       \centering
       \includegraphics[width=0.3\textwidth]{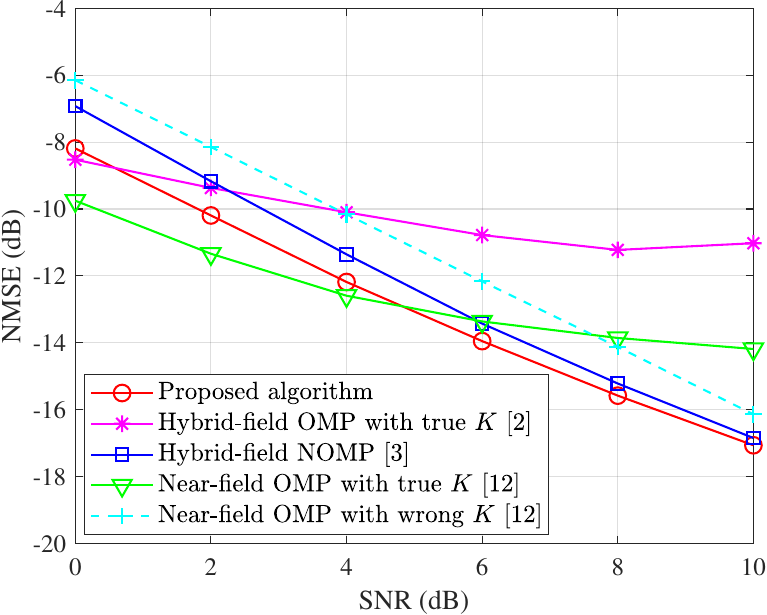}
       \caption{NMSE versus SNR with different algorithms at the current time slot $t>0$, given the historical CSI from the previous time slot $t-1$.}\label{fig:NMSE_SNR}
\end{figure}

We evaluate the channel tracking performance at the current time slot $t>0$, given the historical CSI from the previous time slot $t-1$.
The NMSE performance versus the SNR with different channel estimation/tracking algorithms is depicted in Fig.\,\ref{fig:NMSE_SNR}, in which the hybrid-field OMP \cite{WD-22CL} featuring a composite dictionary to capture mixed-field components and the near-field OMP \cite{CD-22TCOMM} utilizing a specialized polar-domain dictionary with $S = M\nu$ are adopted as benchmarks.
Since both the hybrid-field OMP and the near-field OMP algorithms require prior information of $K$, the results are presented under two cases: the true $K$, i.e., $\hat{K} = K$, and the wrong $K$, i.e., $\hat{K} = 3K$.
It is worth noting that the performance of hybrid-field OMP with wrong $K$ is omitted from Fig.\,\ref{fig:NMSE_SNR}, since it exhibits extremely poor performance, and renders the other comparisons indistinguishable.
It can be observed that the proposed algorithm achieves the best NMSE at the medium SNR regime, verifying its effectiveness.
Besides, compared with the hybrid-field and the near-field OMP algorithms, the proposed algorithm does not necessitate the prior information of $K$, but still achieves superior performance to the hybrid-field OMP algorithm with true $K$ from the low SNR to the medium SNR.
The observed saturation in the high SNR regime for the hybrid-field and near-field OMP algorithms with true $K$ is attributed to the off-the-grid effect. This is because the dominant error source shifts from additive noise to the inherent grid quantization error with the increasing SNR.
Meanwhile, the counter-intuitive performance gain in the wrong $K$ case is attributed to spectral leakage caused by the off-the-grid effect, where the additional grid points help capture the signal energy dispersed across adjacent grids, effectively minimizing the residual error.
Although this mathematically reduces the NMSE, it results in spurious paths that do not reflect the true physical channel geometry.
Furthermore, the proposed algorithm also outperforms the hybrid-field NOMP algorithm \cite{XYang-24WCL} by exploiting the historical CSI.

\begin{figure}[htbp]
       \centering
       \subfloat[]{
       \includegraphics[width=0.22\textwidth]{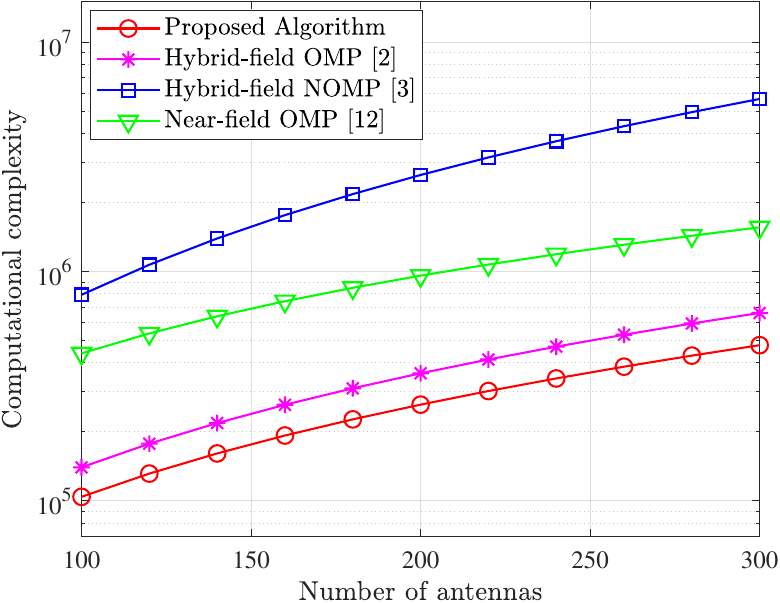}}
       \subfloat[]{
       \includegraphics[width=0.22\textwidth]{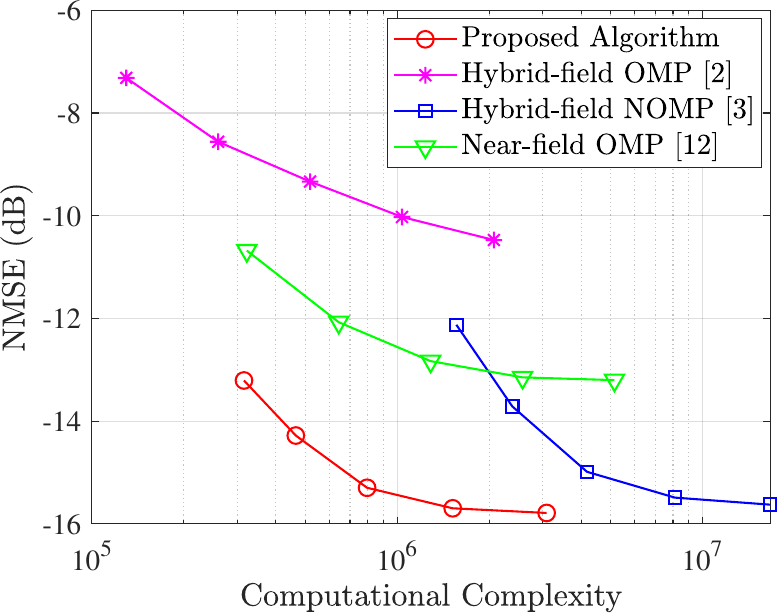}}\\
       \caption{(a) Computational complexity versus the number of antennas with different algorithms. (b) NMSE versus the computational complexity of different algorithms.}\label{fig:complexity}
\end{figure}

To quantify the computational complexity, we summarize the computational complexity of different algorithms in Table\,\ref{table:Complexity} and plot them versus the number of BS antennas $M$ in Fig.\,\ref{fig:complexity}(a).
Since the hybrid-field and the near-field OMP algorithms typically require a large value of $S$, the proposed algorithm exhibits lower complexity than these methods.
Moreover, as the LoS path detection is confined to a narrow search window and the re-detection of the existing NLoS path parameters is unnecessary in the proposed algorithm, its computational complexity is significantly lower than that of the hybrid-field NOMP algorithm.

To further investigate the performance trade-off between the estimation accuracy and the computational complexity, we depict the NMSE performance versus the computational complexity of different algorithms in Fig.\,\ref{fig:complexity}(b) by varying the oversampling factor $\nu$.
For all considered algorithms, the NMSE decreases as the computational complexity increases, which is expected since a larger $\nu$ provides a finer resolution for channel estimation/tracking.
Moreover, it can be observed that the proposed algorithm exhibits the most favorable performance trade-off, achieving superior estimation accuracy with significantly lower computational complexity.

\section{Conclusion}\label{s:Conclusion}
We investigated a hybrid-field channel tracking algorithm in this paper, where the historical CSI was exploited to reduce computational complexity and enhance estimation accuracy.
Firstly, the LoS path was coarsely estimated by performing a sparse peak search within a narrow window, while the coarse estimations of NLoS paths were obtained by ensuring the scatterer survival probability.
Then, a Newton-based algorithm was applied to refine the estimated paths before seeking potential new paths.
Finally, numerical results demonstrated the effectiveness and superiority of the proposed algorithm.

\footnotesize
\bibliographystyle{IEEEtran}

\end{document}